\documentclass[12pt]{article}
\pdfoutput=1

\usepackage{jheppub}
\usepackage[utf8]{inputenc} 
\usepackage{bmpsize}

\usepackage{bm}
\usepackage{amsmath}
\usepackage{eufrak}

\title{Relativistic Hydrodynamic Attractors with Broken Symmetries: Non-Conformal and Non-Homogeneous}

\author{Paul Romatschke$^{1,2}$}
\affiliation{$^1$ Department of Physics, University of Colorado, Boulder, Colorado 80309, USA}
\affiliation{$^2$ Center for Theory of Quantum Matter, University of Colorado, Boulder, Colorado 80309, USA}
\emailAdd{paul.romatschke@colorado.edu}
\abstract{Standard textbooks will state that hydrodynamics requires near-equilibrium to be applicable. Recently, however, out-of-equilibrium attractor solutions for hydrodynamics have been found in kinetic theory and holography in systems with a high degree of symmetry, suggesting the possibility of a genuine out-of-equilibrium formulation of hydrodynamics. This work demonstrates that attractor solutions also occur in non-conformal kinetic theory and spatially non-homogeneous systems, potentially having important implications for the interpretation of experimental data in heavy-ion and proton-proton collisions and relativistic fluid dynamics as a whole. }

\begin{document}
\maketitle
%%%%%%%%%%%%%
%%%%%%%%%%%%%
\section{Introduction}

Recently, attractor solutions for relativistic hydrodynamics have been found in various systems with a high degree of symmetry \cite{Heller:2015dha,Romatschke:2017vte,Spalinski:2017mel,Strickland:2017kux}. Besides their relation to the mathematical theory of resurgence \cite{Buchel:2016cbj,Aniceto:2015mto}, these attractor solutions are interesting because they imply a firm theoretical foundation for the applicability of hydrodynamics in out-of-equilibrium situations \cite{Spalinski:2016fnj,Romatschke:2016hle,Romatschke:2017vte,Florkowski:2017olj}.
A key question in this context is if attractor solutions can be found in systems that do not exhibit additional symmetries. 

A central property of hydrodynamic attractor solutions is that they become indistinguishable from solutions of relativistic dissipative hydrodynamics in the limit of small gradients. As such it is useful to first consider the relativistic generalization of the Navier-Stokes equations\footnote{For the purpose of this discussion, the well-known problems of acausality and instability of the Navier-Stokes equations \cite{Hiscock:1985zz,Hiscock:1983zz} can be safely ignored.} \cite{LL}
\begin{eqnarray}
  D\epsilon +\left(\epsilon+P\right)\nabla_\lambda u^\lambda &=& \frac{\eta}{2}\sigma^{\mu\nu}\sigma_{\mu\nu}+\zeta \left(\nabla_\lambda u^\lambda\right)^2\,,\\
  \left(\epsilon+P\right)D u^\alpha+\nabla_\perp^\alpha P&=& \Delta_\nu^\alpha \nabla_\mu \left(\eta \sigma^{\mu\nu}+\zeta \Delta^{\mu\nu} \nabla_\lambda u^\lambda\right)\,,
\end{eqnarray}
where $\epsilon,P,u^\mu$ are the fluid's local energy density, pressure and velocity, and $\nabla_\mu$ is the geometric covariant derivative. Furthermore, $\eta,\zeta$ are the shear and bulk viscosity coefficients, $D\equiv u^\mu \nabla_\mu$, $\nabla_\perp^\mu\equiv \Delta^{\mu\nu}\nabla_\nu$, $\Delta^{\mu\nu}\equiv g^{\mu\nu}+u^\mu u^\nu$, $\sigma^{\mu\nu}\equiv \nabla_\perp^\mu u^\nu+\nabla_\perp^\nu u^\mu-\frac{2}{3}\Delta^{\mu\nu} \nabla_\mu u^\mu$ and the mostly plus convention for the metric tensor $g_{\mu\nu}$ will be used.

In the absence of any conserved charges, the Navier-Stokes equation for the energy-density evolution may be rewritten as
\begin{equation}
  \label{eq:scal1}
  \frac{D\epsilon}{(\epsilon+P)\nabla_\lambda u^\lambda}=\frac{D \ln s}{\nabla_\lambda u^\lambda}=-1+\frac{\eta}{2 s}\frac{\sigma^{\mu\nu}\sigma_{\mu\nu}}{T \nabla_\lambda u^\lambda}+\frac{\zeta}{s } \frac{\nabla_\lambda u^\lambda}{T}\,,
\end{equation}
where $s,T$ are the entropy density and temperature, respectively, related to $\epsilon,P$ via the usual thermodynamic relations. For relativistic systems, which are never incompressible, the expansion scalar $\nabla_\lambda u^\lambda$ is generally non-vanishing unless global equilibrium is reached. Eq.~(\ref{eq:scal1}) then implies that for a given system, the time-evolution of the quantity
\begin{equation}
\label{eq:att}
  A_1=\frac{D \ln s}{\nabla_\lambda u^\lambda}\,,
\end{equation}
will behave similarly for small gradients irrespective of initial conditions. As an example, consider systems exhibiting conformal symmetry $\zeta=0$ and $\frac{\eta}{s}={\rm const}$. In this case, Eq.~(\ref{eq:scal1}) implies that the time evolution of $\frac{D \ln s}{\nabla_\lambda u^\lambda}$ for arbitrary initial conditions will collapse onto a single curve when expressed as a function of 
$\frac{2 s T \nabla_\lambda u^\lambda}{\eta \sigma^{\mu\nu}\sigma_{\mu\nu}}$.

As another example, consider non-conformal systems where the trace of the energy-momentum tensor of the Navier-Stokes equation is given by ${\rm Tr}\, T^{\mu\nu}=-\epsilon+3 P-3 \zeta \nabla_\lambda u^\lambda$. The trace of the energy-momentum tensor corresponds to the sum of temporal and spatial eigenvalues, and thus encodes information about the effective equation of state the system is experiencing. In equilibrium, $\nabla_\lambda u^\lambda=0$, and hence the trace anomaly implies an equilibrium equation of state. For non-conformal systems out of equilibrium, the Navier-Stokes equation implies that  the time-evolution of the quantity
\begin{equation}
  A_2=\frac{{\rm Tr}\,T^{\mu\nu}+\epsilon-3 P}{\zeta T}\,,
\end{equation}
will behave similarly for small gradients irrespective of initial conditions if expressed as a function of the inverse gradient strength 
\begin{equation}
\label{eq:var}
\Gamma\equiv 
\left[\frac{\eta}{2 s}\frac{\sigma^{\mu\nu}\sigma_{\mu\nu}}{T \nabla_\lambda u^\lambda}+\frac{\zeta}{s } \frac{\nabla_\lambda u^\lambda}{T}\right]^{-1}\,.
\end{equation}
Clearly, $\Gamma$ reduces to the expression for the scaling strength found in the conformal case $\zeta=0$ above.

Real systems exhibit deviations from the behavior predicted by the Navier-Stokes equation at finite gradient strength. Nevertheless, solutions will tend to the Navier-Stokes solution in the limit of small gradients, such that it acts as an attractor solution. Less trivial is the question of whether such an attractor solution extends to the regime of moderate or even large gradients.
In practice, one can search for attractor solutions in real systems or microscopic theories by e.g. studying the time-evolution of quantities such as $A_1,A_2$. For $A_1$, this has been successfully achieved in system with conformal symmetry and restricted spatial dynamics (Bjorken flow in the longitudinal direction\cite{Bjorken:1982qr} and spatially homogeneous in the transverse directions) in Refs.~\cite{Heller:2015dha,Romatschke:2017vte,Spalinski:2017mel,Strickland:2017kux}.
The present work is trying to extend the understanding of hydrodynamic attractor solutions by studying non-conformal systems and systems that allow for spatial dynamics.

\section{Non-Conformal Attractor in Kinetic Theory}

Consider a gas of particles with mass $m$ undergoing one-dimensional boost-invariant expansion according to Bjorken \cite{Bjorken:1982qr}. It is convenient to work in Milne coordinates proper time $\tau=\sqrt{t^2-z^2}$ and spacetime rapidity ${\rm arctanh}(z/t)$ for this system. Kinetic theory in the relaxation-time approximation is defined by a single-particle on-shell distribution function $f(x^\mu,p^\mu)$ which obeys the Boltzmann BGK equation \cite{Boltzmann:1872,PhysRev.94.511}
\begin{equation}
\label{eq:boltzmann}
  p^\mu \partial_\mu f-p^\lambda p^\sigma \Gamma^\mu_{\lambda \sigma} \frac{\partial f}{\partial p^\mu}=\frac{p^\mu u_\mu \left(f-f_{\rm eq}\right)}{\tau_R}\,,
  \end{equation}
where $\Gamma^\mu_{\lambda \sigma}$ are the Christoffel symbols for Milne coordinates and $f_{\rm eq}=2\pi^2 e^{p^\mu u_\mu/T}$. In these expressions, $u^\mu(x^\mu),T(x^\mu)$ are related to the time-like eigenvector and eigenvalue of the energy-momentum tensor $T^{\mu\nu}(x^\mu)=\int d\chi p^\mu p^\nu f$ as\footnote{Note that the integration measure is given by
$d\chi\equiv\frac{d^4 p}{(2\pi)^4} \sqrt{-{\rm det}g_{\mu \nu}}\, 2\Theta(p^0) (2\pi) \delta\left(-g_{\mu\nu} p^\mu p^\nu-m^2\right)$.}
\begin{equation}
  \label{eq:edef}
  \epsilon u^\mu \equiv - T^{\mu\nu} u_\nu\,,
\end{equation}
with the normalization condition $u^\mu u_\mu=-1$. Here $\epsilon$ can be recognized as the local energy density. For a massive gas at temperature $T$ in equilibrium, the relation between energy density and temperature is readily calculated from (\ref{eq:edef}) with $f=f_{\rm eq}$. Working in units where the particle mass $m=1$, one finds \cite{Romatschke:2011qp}
\begin{eqnarray}
  \label{eq:thermo}
  &\epsilon(T)=3 T^2 K_{2}\left(T^{-1}\right)+ T K_{1}\left(T^{-1}\right), \quad P(T)=T^2 K_{2}\left(T^{-1}\right)&\\
  &  s(T)=K_3\left(T^{-1}\right)\,,\quad c_s^2(T)=\left(3+T^{-1}\frac{K_3\left(T^{-1}\right)}{K_2\left(T^{-1}\right)}\right)^{-1}&\\
  &\eta(T)=\frac{\tau_R}{T}\int_0^T dT^\prime T^\prime s(T^\prime)\,,\quad
  \zeta(T)=\frac{5}{3}\eta(T)- \tau_R T s(T)c_s^2(T)\, &
\end{eqnarray}
for the energy density, pressure $P$, entropy density $s$, speed of sound squared $c_s^2$ as well as shear and bulk viscosity coefficients, respectively. Here $K(x)$ denote modified Bessel functions.

Out of equilibrium, there is no temperature $T$ for the system, but there always is an energy density that can be found from Eq.~(\ref{eq:edef}). To find the parameter $T(\epsilon)$ in $f_{\rm eq}$, note that integration $\int d\chi p^\nu$ of Eq.~(\ref{eq:boltzmann}) leads to covariant conservation of the energy-momentum tensor iff $u_\mu T^{\mu\nu}=u_\mu T^{\mu\nu}_{\rm eq}$ \cite{Romatschke:2011qp}. Therefore, $T(\epsilon)$ out of equilibrium is required to be chosen such that the energy density in (\ref{eq:thermo}) matches the time-like eigenvalue $\epsilon$ of $T^{\mu\nu}$. Note that this is \textit{not} implying that the system evolves with an equilibrium equation of state, which is a relation between the time-like and space-like eigenvalues of $T^{\mu\nu}$, but rather only fixing the setup of the equation (\ref{eq:boltzmann}).

In a system that is homogeneous with respect to transverse coordinates ${\bf x}_\perp=(x,y)$ and boost-invariant (independent of space-time rapidity), Eq.~(\ref{eq:boltzmann}) implies the following integral equation for the evolution of the energy-density as a function of proper time \cite{Florkowski:2014sfa}:
\begin{eqnarray}
  \label{eq:eevol}
  \epsilon\left(T(\tau)\right)&=&\Lambda_0^4 D(\tau,\tau_0) h_2\left(\frac{\tau_0}{\tau\sqrt{1+\xi_0}},\Lambda_0^{-1}\right)+\int_{\tau_0}^{\tau}\frac{d\tau^\prime}{\tau_R}D(\tau,\tau^\prime) h_2\left(\frac{\tau^\prime}{\tau},T^{-1}(\tau^\prime)\right)\,,\\
  h_2(y,z)&=&\frac{y}{2}\int_0^\infty du u^3 e^{-\sqrt{u^2+z^2}}\left(\sqrt{y^2+z^2/u^2}+\frac{1+z^2/u^2}{\sqrt{y^2-1}} {\rm arctanh}\sqrt{\frac{y^2-1}{y^2+z^2/u^2}}\right)\,,\nonumber\\
  D(\tau_2,\tau_1)&=&e^{-\int_{\tau_1}^{\tau_2} \frac{d\tau^\prime}{\tau_R}}\,.\nonumber
\end{eqnarray}
For simplicity, $\tau_R=C_\pi/T$ with constant $C_\pi$ was chosen in the following. Initial conditions for the system are characterized by choosing a value of $\xi_0\in[-1,\infty)$ and $\Lambda_0$ at $\tau=\tau_0$. Numerical solutions to Eq.~(\ref{eq:eevol}) for $\Lambda_0=1$ and various values of $\xi_0$ can be generated by the methods outlined in Ref.~\cite{Florkowski:2013lza}. For a spatially homogeneous and boost-invariant system, Eq.~(\ref{eq:att}) becomes $A_1(\tau)=\tau \partial_\tau \ln s$ and $\frac{T \nabla_\lambda u^\lambda}{\sigma^{\mu\nu}\sigma_{\mu\nu}}=\frac{8}{3}\tau T$, $\frac{\nabla_\lambda u^\lambda}{T}=\tau T$ lead to $\Gamma=\tau/\gamma_s$ with $\gamma_s\equiv \frac{4}{3}\frac{\eta}{\epsilon+P}+\frac{\zeta}{\epsilon+P}$ the temperature-dependent sound attenuation length.

\begin{figure}[t]
  \includegraphics[width=0.49\linewidth]{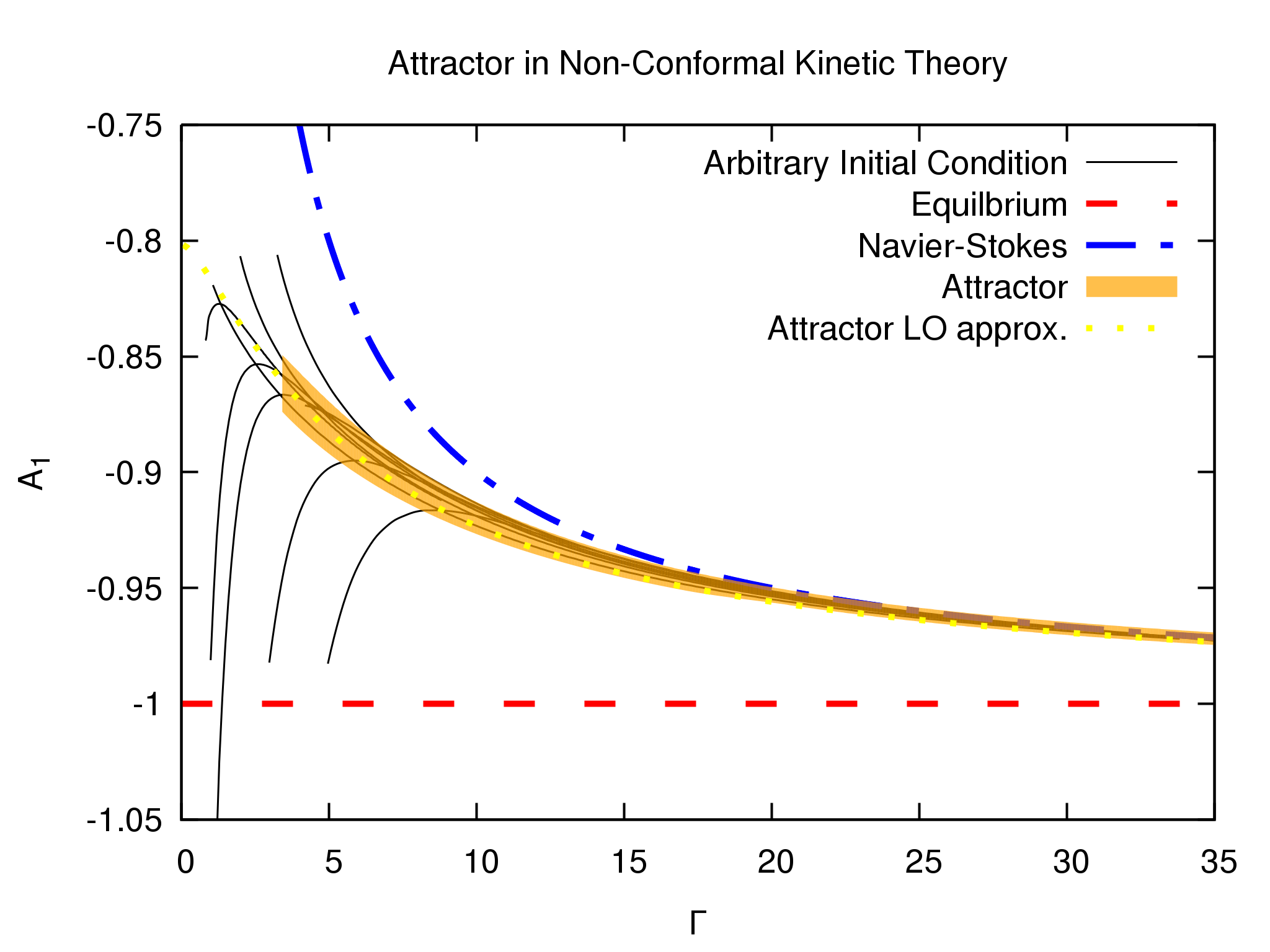}
  \hfill
   \includegraphics[width=0.49\linewidth]{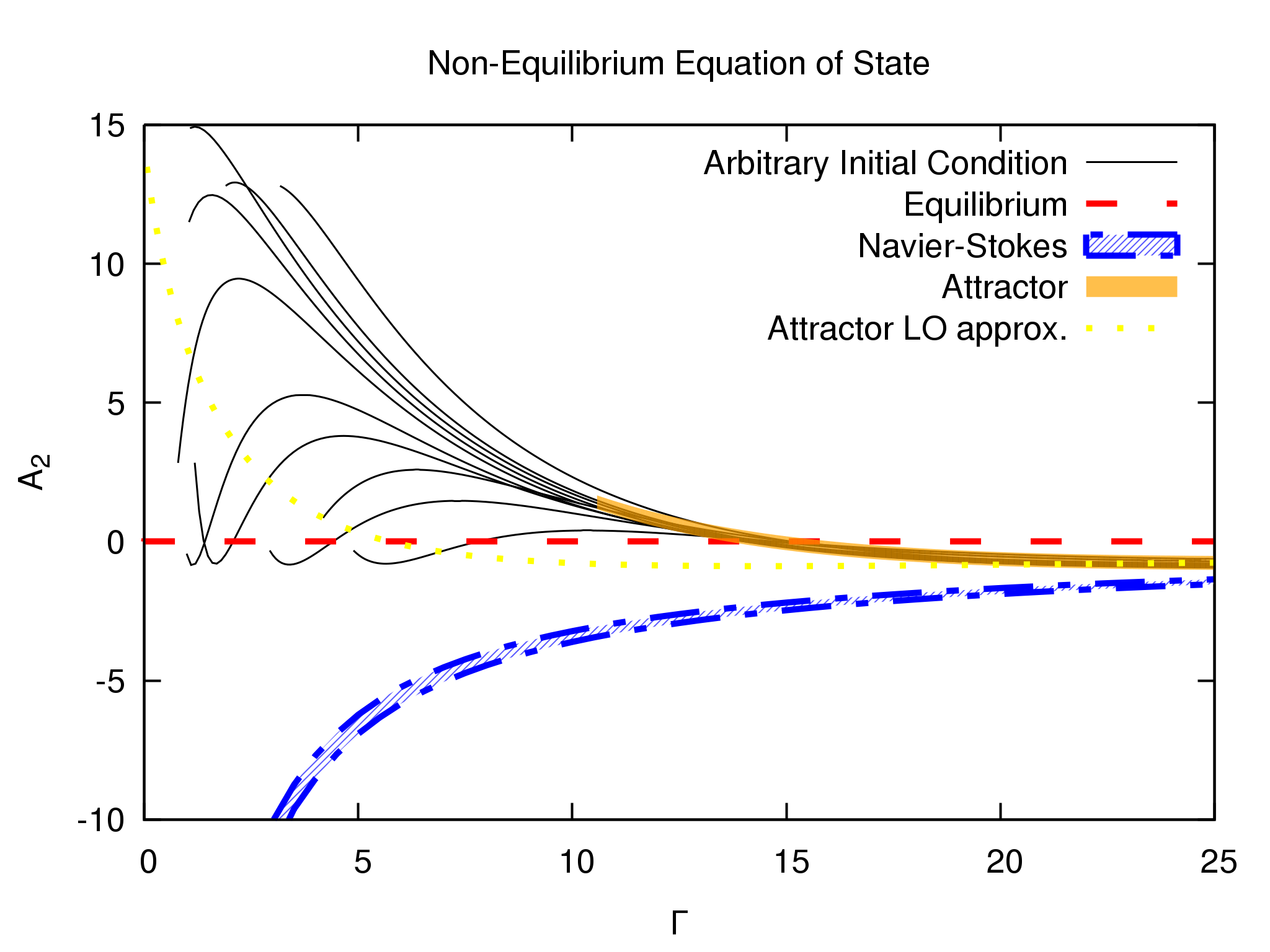}
  \caption{\label{fig:nc} Scaling variables $A_1,A_2$ for non-conformal kinetic theory in Bjorken flow with $\Lambda_0=1$, $\tau_R=0.1/T$ as a function of inverse gradient strength $\Gamma$. Note that for $A_2$, scaling the gradient strength with the sound attenuation length implies that the Navier-Stokes result corresponds to an area rather than a single curve. The curves labeled 'Attractor LO approx.' are generated by solving Eq.~(\ref{eq:condition}) for $\xi_0(\tau_0)$ and evaluating $A_1,A_2$ using (\ref{eq:stuff}). }
\end{figure}

The results from numerically solving Eq.~(\ref{eq:eevol}) are shown in 
Fig.~\ref{fig:nc}, along with the results from the Navier-Stokes equation. One observes that for arbitrary initial choices of $\xi_0$ at fixed $\tau_0$, the subsequent evolution tend to cluster in special 'attractor solutions' which eventually merge with the Navier-Stokes results. In the case of $A_1$, it is possible to find points close to the attractor solution by employing the technique outlined in \cite{Romatschke:2017vte}, namely the 'slow-roll' condition $\left.\partial_\tau A_1\right|_{\tau=\tau_0,\xi=\xi_0}=0$ \cite{Heller:2015dha}. For the case at hand, this condition becomes
  \begin{equation}
    \label{eq:condition}
    \left.\frac{\left(\epsilon+P\right)\left(\partial_\tau \epsilon+\tau \partial_\tau^2 \epsilon\right)-\tau \left(\partial_\tau \epsilon\right)^2\left(1+c_s^2\right)}{\left(\epsilon+P\right)^2}\right|_{\tau=\tau_0,\xi=\xi_0}=0\,,
  \end{equation}
  where\footnote{Another useful relation is $\epsilon(T)=T^4 h_2(1,T^{-1})$.}
  \begin{eqnarray}
    \label{eq:stuff}
&    \epsilon(\tau_0)=\Lambda_0^4 h_2\left(\frac{1}{\sqrt{1+\xi_0}},\Lambda_0^{-1}\right)\,,\quad
    \partial_\tau\epsilon(\tau_0)=-\frac{\Lambda_0^4}{\tau_0 \sqrt{1+\xi_0}} h_2^{(1,0)}\left(\frac{1}{\sqrt{1+\xi_0}},\Lambda_0^{-1}\right)\,,&\\
  &  \partial_\tau^2 \epsilon(\tau_0)=\frac{-1}{\tau_R}\left(\partial_\tau \epsilon(\tau_0)+\frac{(\epsilon+P)}{\tau_0}\right)+\Lambda_0^4\left(\frac{h_2^{(2,0)}\left(\frac{1}{\sqrt{1+\xi_0}},\Lambda_0^{-1}\right)}{\tau_0^2(1+\xi_0)}+\frac{2h_2^{(1,0)}\left(\frac{1}{\sqrt{1+\xi_0}},\Lambda_0^{-1}\right)}{\tau_0^2\sqrt{1+\xi_0}} \right)\,.\nonumber&
  \end{eqnarray}
  Solving (\ref{eq:condition}) numerically for $\tau_R=\frac{0.4}{T}$ and $\tau_0=0.1$, $\Lambda_0=1$ (all in mass units $m=1$) leads to $\xi_0\simeq 17.8$. Using this value of $\xi_0$ as initial condition, an attractor solution for $A_1$ may be constructed numerically by solving Eq.~(\ref{eq:eevol}). Unlike the case of conformal theories, repeating the above procedure for different starting times $t_0$ will lead to a slightly different attractor curve. This can be understood from the fact that, in the non-conformal case, $A_1$ is not a simple function of $\tau T$ alone because there is an additional mass scale to contend with. As a consequence, an envelope of attractor curves for $A_1$ is shown in Fig.~\ref{fig:nc}, suggesting that the attractor in the non-conformal case is an extended object. The width of the attractor envelope is related to the value of $\Lambda_0^{-1}$  and I have checked that the conformal (zero-width) attractor from Ref.~\cite{Romatschke:2017vte} is recovered in the limit $\Lambda_0\rightarrow \infty$.

  Results for $A_2$ are also shown in Fig.~\ref{fig:nc}. One observes a clustering of trajectories for arbitrary initial condition similar to $A_2$, again suggesting an attractor solution at early times that is distinct from the Navier-Stokes result. However, the approach of individual trajectories to the $A_2$ attractor solution seems to be slower than for $A_1$. The region labeled 'Attractor' in Fig.~\ref{fig:nc} marks the area where different $A_1$ attractor solutions to Eq.~(\ref{eq:eevol}) have merged. This demonstrates that there are special solutions to Eq.~(\ref{eq:eevol}) which are attractors for $A_1$ and $A_2$ simultaneously.

  \section{Non-Homogeneous Attractor in rBRSSS}

  All attractor solutions discussed so far where restricted to spatially homogeneous systems, begging the question if attractor solutions survive if the spatial dynamics is not strongly restricted. To study this question, consider the mock-microscopic theory of resummed BRSSS (rBRSSS for short) in conformal symmetry, which is defined by an energy momentum tensor $T^{\mu\nu}=\epsilon u^\mu u^\nu+P \Delta^{\mu\nu}+\pi^{\mu\nu}$ with the dynamic shear stress $\pi^{\mu\nu}$ obeying the equations of motion \cite{Baier:2007ix}
  \begin{eqnarray}
   \label{eq:rBRSSS}
   \pi^{\mu \nu}&=&-\eta \sigma^{\mu \nu}-\tau_\pi \left[\left< D \pi^{\mu \nu}\right>+\frac{4}{3} \pi^{\mu \nu} \nabla_\lambda^\perp u^\lambda\right]
+\kappa\left[R^{<\mu \nu>}-2 u_\lambda u_\rho R^{\lambda <\mu \nu>\rho}\right]
\nonumber\\
&&+\frac{\lambda_1}{\eta^2} \pi^{<\mu}_{\quad \lambda}\pi^{\nu> \lambda}-\frac{\lambda_2}{\eta} \pi^{<\mu}_{\quad \lambda} \Omega^{\nu> \lambda}
+\lambda_3 \Omega^{<\mu}_{\quad \lambda}\Omega^{\nu> \lambda}\,.
 \end{eqnarray}
  In the following, only flat-space systems where the Ricci and Riemann tensors vanish are considered. Also, for simplicity I will set $\lambda_1=\lambda_2=\lambda_3=0$. The rBRSSS equations of motion are causal as long as $\tau_\pi\geq \frac{2 \eta}{s T}$, cf. Ref.~\cite{Romatschke:2009im}.

 \begin{figure}[t]
\center
      \includegraphics[width=0.7\linewidth]{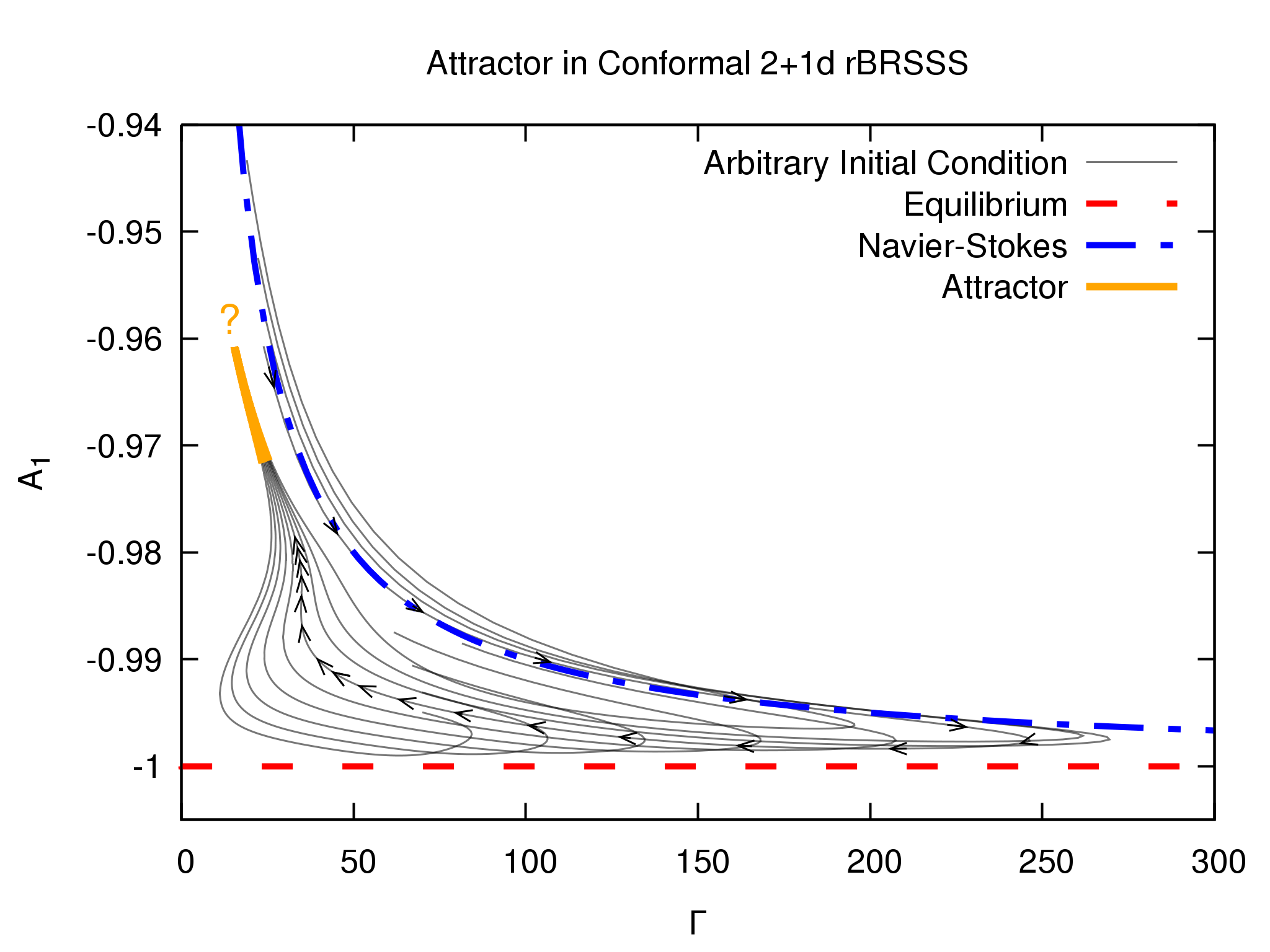}
      \caption{\label{fig2} Selected trajectories of $A_1(\Gamma)$ from solving rBRSSS equations numerically in 2+1d for a Au+Au collision at $b=8$ fm impact parameter with $\frac{\eta}{s}=0.16,\tau_\pi T=\frac{4 \eta}{s}$. The direction of the time evolution is indicated by arrows on one trajectory. While at early times (not shown), trajectories are far separated and strongly dependent on initial conditions, at late times one observes a clustering of trajectories near an apparent attractor solution. The attractor solution does not stop near $\Gamma_1\simeq 16, A_1\simeq -0.96$, but system evolution of $A_1(\Gamma)$ slows down dramatically in real time near this point, making it computationally expensive to continue tracking the attractor.}
\end{figure}
  
  For a spatially homogeneous system, the rBRSSS equations have been shown to possess a hydrodynamic attractor solution \cite{Heller:2015dha}. Fortunately, numerical solvers for rBRSSS equations are readily available for spatially non-homogeneous systems \cite{Romatschke:2007mq,Dusling:2007gi,Chaudhuri:2007qp,Song:2008si,Schenke:2010rr}. I will be using the VH2+1 solver from Ref.~\cite{Romatschke:2007mq}, which solves the rBRSSS equations for systems that are boost-invariant, but otherwise unrestricted in terms of the dynamics in transverse coordinates ${\bf x}_\perp=(x,y)$. Using an optical Glauber model of a Au+Au collision at an impact parameter of $8$ fm with AdS/CFT pre-equilibrium flow \cite{vanderSchee:2013pia} as initial condition\footnote{For the details of the implementation of the Glauber model see for instance Ref.~\cite{Romatschke:2009im}. In essence, the initial conditions considered here are qualitatively similar to a two-dimensional Gaussian energy density $\epsilon(\tau=\tau_0,x,y)$ with different width in $x,y$.}$^{,}$\footnote{While it would have been possible to consider simpler initial conditions, the choice of Glauber+pre-equilibrium flow corresponds to studying attractors in the superSONIC model used in relativistic ion collision phenomenology \cite{Romatschke:2015gxa,Weller:2017tsr}.}, the equations of motion are solved numerically on a lattice in transverse space ${\bf x}_\perp$. On each lattice point at each time-step, it is possible to evaluate $A_1,\Gamma$ by locally calculating (\ref{eq:att}),(\ref{eq:var}) numerically. A representative plot of the resulting trajectories is shown in Fig.~\ref{fig2}.

As can be seen from this figure, trajectories in the $A_1,\Gamma$ plane are initially far separated, with some of the trajectories being close to the Navier-Stokes result, while others are not. However, at late times in the system evolution when gradients are no longer dominated solely by longitudinal Bjorken flow, trajectories cluster near an apparent attractor solution. Once the system comes close to the regime near $\Gamma\simeq 16, A_1\simeq -0.96$, evolution of $A_1(\Gamma)$ slows down dramatically in real time, making it computationally expensive to continue tracking the attractor. I have checked that the apparent attractor remains unaffected by choosing different initial temperatures, starting times and impact parameters for the rBRSSS solution.

\section{Findings and Interpretations}

Prior to the present study, relativistic hydrodynamic attractor solutions had been identified in systems with a high degree of symmetry (conformal symmetry and spatial homogeneity). In this work, the existence of attractor solutions for system with broken symmetries (non-conformal and spatially non-homogeneous) was investigated.

Findings:
\begin{itemize}
  \item
    In the case of non-conformal kinetic theory undergoing Bjorken flow, there exists an attractor solution for the quantity $A_1$ in Eq.~(\ref{eq:att}) which is qualitatively similar to, but quantitatively different from, the known attractor solution of the conformal case \cite{Romatschke:2017vte}.
  \item
    In the case of non-conformal kinetic theory undergoing Bjorken flow, points close to the non-conformal attractor can be calculated by a 'slow-roll' approximation (\ref{eq:condition}).
  \item
    In the case of non-conformal kinetic theory undergoing Bjorken flow, the non-conformal attractor solution for $A_1$ also acts as an attractor for the quantity $A_2$, which controls the non-equilibrium equation of state (the relation between energy and non-equilibrium pressure).
  \item
    In the case of conformal non-homogeneous rBRSSS theory, there exists an attractor solution for the quantity $A_1(\Gamma)$ with definitions (\ref{eq:att}), (\ref{eq:var}) that can be constructed numerically.
  \item
    In the case of conformal non-homogeneous rBRSSS theory, the attractor solution $A_1(\Gamma)$ is only partially known because the numerical evolution slows down dramatically.
\end{itemize}

Interpretations:
\begin{itemize}
\item
  Together with previous results on this subject, the present work strongly suggests that non-analytic attractor solutions for relativistic hydrodynamics exist in a broad class of theories regardless of the underlying symmetries.
\item
  Traditionally, relativistic hydrodynamics has been defined via a gradient expansion, with Euler equation, Navier-Stokes, BRSSS \cite{Baier:2007ix,Romatschke:2009kr} and Grozdanov-Kaplis theory \cite{Grozdanov:2015kqa} the respective complete $0^{\rm th}$, $1^{\rm st}$, $2^{\rm nd}$ and $3^{\rm rd}$ order realizations. However, the hydrodynamic gradient series is expected to be divergent \cite{Heller:2013fn,Heller:2015dha,Aniceto:2015mto,Buchel:2016cbj,Florkowski:2016zsi}, calling into question the meaning of solutions to the hydrodynamic gradient series for any non-vanishing gradient strength \cite{Romatschke:2016hle}. The existence of hydrodynamic attractor solutions provides this meaning and serves as the foundation for a new, yet to be elaborated, theory of hydrodynamics out-of-equilibrium \cite{Romatschke:2017vte,Florkowski:2017olj}.
\item
  The existence of an apparent attractor for the non-equilibrium equation of state in Fig.~\ref{fig:nc}b eliminates the last remnant of the standard textbook 'hydrodynamics requires equilibrium' paradigm. While the relation between $\epsilon,P$ is by construction controlled by the thermodynamic results (\ref{eq:thermo}), the system experiences a non-equilibrium pressure $P_{\rm eff}=P-\zeta_B(\epsilon,\nabla) \nabla_\lambda u^\lambda$ where $\zeta_B$ is a non-analytic function that depends on both the energy density as well as the gradient strength (denoted formally as $\nabla$), similar to the shear viscosity coefficient $\eta_B$ defined in Ref.~\cite{Romatschke:2017vte}.  Though not interpreted in this fashion, non-equilibrium equations of state are now routinely used (and indeed required!) to provide precision fits of hydrodynamic models of relativistic ion collisions to experimental data \cite{Ryu:2015vwa,Habich:2015rtj}.
\item
Given the above findings, it can be considered reasonably likely that an attractor solution for QCD exists in the context of relativistic ion collisions. This attractor solution would result in 'hydrodynamic-like' behavior of the system without any requirement of system equilibration. Therefore, it would naturally explain the experimentally observed 'hydrodynamic-like' signatures in relativistic heavy-ion and proton-proton collisions \cite{Schukraft:2017nbn} and possibly even indicate hydrodynamic behavior in electron-positron collisions \cite{Nagle:2017sjv}.
\item
Besides the immediate application to the field of relativistic nuclear collisions, the existence of hydrodynamic attractor solutions may have important implications for relativistic fluid dynamics in general, e.g. by providing a firm foundation for viscous cosmologies \cite{Zimdahl:1996ka,Gagnon:2011id,Blas:2015tla}.
  \end{itemize}

\section*{Note Added in Proof}

While this work was being reviewed, results for the non-conformal kinetic theory attractor were presented by a different group in Ref.~\cite{Florkowski:2017jnz}. The results in Ref.~\cite{Florkowski:2017jnz} appear to be in full agreement with this work.

\section{Acknowledgments}

I'd like to thank S.~Jeon, J.~Casalderrey-Solana and M.~Spali\'nksi for fruitful discussions on this topic and the organizers for the Initial Stages 2017 conference for providing such a nice and stimulating conference. Furthermore, I'd like to thank F.~Becattini and S.~Mrowczynski for challenging me to demonstrate the existence of attractor solutions beyond highly symmetric systems. This research was funded by the Department of Energy, award number DE-SC0017905.

\bibliographystyle{JHEP}
\bibliography{pp-hydro}

\providecommand{\href}[2]{#2}\begingroup\raggedright\begin{thebibliography}{10}

\bibitem{Heller:2015dha}
M.~P. Heller and M.~Spaliński, {\it {Hydrodynamics Beyond the Gradient
  Expansion: Resurgence and Resummation}},  {\em Phys. Rev. Lett.} {\bf 115}
  (2015), no.~7 072501, [\href{http://arxiv.org/abs/1503.07514}{{\tt
  arXiv:1503.07514}}].

\bibitem{Romatschke:2017vte}
P.~Romatschke, {\it {Far From Equilibrium Fluid Dynamics}},
  \href{http://arxiv.org/abs/1704.08699}{{\tt arXiv:1704.08699}}.

\bibitem{Spalinski:2017mel}
M.~Spaliński, {\it {On the hydrodynamic attractor of Yang-Mills plasma}},
  \href{http://arxiv.org/abs/1708.01921}{{\tt arXiv:1708.01921}}.

\bibitem{Strickland:2017kux}
M.~Strickland, J.~Noronha, and G.~Denicol, {\it {The anisotropic
  non-equilibrium hydrodynamic attractor}},
  \href{http://arxiv.org/abs/1709.06644}{{\tt arXiv:1709.06644}}.

\bibitem{Buchel:2016cbj}
A.~Buchel, M.~P. Heller, and J.~Noronha, {\it {Entropy Production,
  Hydrodynamics, and Resurgence in the Primordial Quark-Gluon Plasma from
  Holography}},  {\em Phys. Rev.} {\bf D94} (2016), no.~10 106011,
  [\href{http://arxiv.org/abs/1603.05344}{{\tt arXiv:1603.05344}}].

\bibitem{Aniceto:2015mto}
I.~Aniceto and M.~Spaliński, {\it {Resurgence in Extended Hydrodynamics}},
  {\em Phys. Rev.} {\bf D93} (2016), no.~8 085008,
  [\href{http://arxiv.org/abs/1511.06358}{{\tt arXiv:1511.06358}}].

\bibitem{Spalinski:2016fnj}
M.~Spaliński, {\it {Small systems and regulator dependence in relativistic
  hydrodynamics}},  {\em Phys. Rev.} {\bf D94} (2016), no.~8 085002,
  [\href{http://arxiv.org/abs/1607.06381}{{\tt arXiv:1607.06381}}].

\bibitem{Romatschke:2016hle}
P.~Romatschke, {\it {Do nuclear collisions create a locally equilibrated
  quark-gluon plasma?}},  {\em Eur. Phys. J.} {\bf C77} (2017), no.~1 21,
  [\href{http://arxiv.org/abs/1609.02820}{{\tt arXiv:1609.02820}}].

\bibitem{Florkowski:2017olj}
W.~Florkowski, M.~P. Heller, and M.~Spaliński, {\it {New theories of
  relativistic hydrodynamics in the LHC era}},
  \href{http://arxiv.org/abs/1707.02282}{{\tt arXiv:1707.02282}}.

\bibitem{Hiscock:1985zz}
W.~A. Hiscock and L.~Lindblom, {\it {Generic instabilities in first-order
  dissipative relativistic fluid theories}},  {\em Phys. Rev.} {\bf D31} (1985)
  725--733.

\bibitem{Hiscock:1983zz}
W.~A. Hiscock and L.~Lindblom, {\it {Stability and causality in dissipative
  relativistic fluids}},  {\em Annals Phys.} {\bf 151} (1983) 466--496.

\bibitem{LL}
L.~Landau and E.~Lifshitz, {\it {Fluid Mechanics}},  {\em Elsevier} {\bf 2nd
  edition} (1987).

\bibitem{Bjorken:1982qr}
J.~D. Bjorken, {\it {Highly Relativistic Nucleus-Nucleus Collisions: The
  Central Rapidity Region}},  {\em Phys. Rev.} {\bf D27} (1983) 140--151.

\bibitem{Boltzmann:1872}
L.~Boltzmann, {\em Vorlesungen \"uber Gastheorie}.
\newblock Johann Ambrosius Barth Verlag, 1896.

\bibitem{PhysRev.94.511}
P.~L. Bhatnagar, E.~P. Gross, and M.~Krook, {\it A model for collision
  processes in gases. i. small amplitude processes in charged and neutral
  one-component systems},  {\em Phys. Rev.} {\bf 94} (May, 1954) 511--525.

\bibitem{Romatschke:2011qp}
P.~Romatschke, {\it {Relativistic (Lattice) Boltzmann Equation with Non-Ideal
  Equation of State}},  {\em Phys. Rev.} {\bf D85} (2012) 065012,
  [\href{http://arxiv.org/abs/1108.5561}{{\tt arXiv:1108.5561}}].

\bibitem{Florkowski:2014sfa}
W.~Florkowski, E.~Maksymiuk, R.~Ryblewski, and M.~Strickland, {\it {Exact
  solution of the (0+1)-dimensional Boltzmann equation for a massive gas}},
  {\em Phys. Rev.} {\bf C89} (2014), no.~5 054908,
  [\href{http://arxiv.org/abs/1402.7348}{{\tt arXiv:1402.7348}}].

\bibitem{Florkowski:2013lza}
W.~Florkowski, R.~Ryblewski, and M.~Strickland, {\it {Anisotropic Hydrodynamics
  for Rapidly Expanding Systems}},  {\em Nucl. Phys.} {\bf A916} (2013)
  249--259, [\href{http://arxiv.org/abs/1304.0665}{{\tt arXiv:1304.0665}}].

\bibitem{Baier:2007ix}
R.~Baier, P.~Romatschke, D.~T. Son, A.~O. Starinets, and M.~A. Stephanov, {\it
  {Relativistic viscous hydrodynamics, conformal invariance, and holography}},
  {\em JHEP} {\bf 04} (2008) 100, [\href{http://arxiv.org/abs/0712.2451}{{\tt
  arXiv:0712.2451}}].

\bibitem{Romatschke:2009im}
P.~Romatschke, {\it {New Developments in Relativistic Viscous Hydrodynamics}},
  {\em Int. J. Mod. Phys.} {\bf E19} (2010) 1--53,
  [\href{http://arxiv.org/abs/0902.3663}{{\tt arXiv:0902.3663}}].

\bibitem{Romatschke:2007mq}
P.~Romatschke and U.~Romatschke, {\it {Viscosity Information from Relativistic
  Nuclear Collisions: How Perfect is the Fluid Observed at RHIC?}},  {\em Phys.
  Rev. Lett.} {\bf 99} (2007) 172301,
  [\href{http://arxiv.org/abs/0706.1522}{{\tt arXiv:0706.1522}}].

\bibitem{Dusling:2007gi}
K.~Dusling and D.~Teaney, {\it {Simulating elliptic flow with viscous
  hydrodynamics}},  {\em Phys. Rev.} {\bf C77} (2008) 034905,
  [\href{http://arxiv.org/abs/0710.5932}{{\tt arXiv:0710.5932}}].

\bibitem{Chaudhuri:2007qp}
A.~K. Chaudhuri, {\it {Saturation of elliptic flow and shear viscosity}},
  \href{http://arxiv.org/abs/0708.1252}{{\tt arXiv:0708.1252}}.

\bibitem{Song:2008si}
H.~Song and U.~W. Heinz, {\it {Multiplicity scaling in ideal and viscous
  hydrodynamics}},  {\em Phys. Rev.} {\bf C78} (2008) 024902,
  [\href{http://arxiv.org/abs/0805.1756}{{\tt arXiv:0805.1756}}].

\bibitem{Schenke:2010rr}
B.~Schenke, S.~Jeon, and C.~Gale, {\it {Elliptic and triangular flow in
  event-by-event (3+1)D viscous hydrodynamics}},  {\em Phys. Rev. Lett.} {\bf
  106} (2011) 042301, [\href{http://arxiv.org/abs/1009.3244}{{\tt
  arXiv:1009.3244}}].

\bibitem{vanderSchee:2013pia}
W.~van~der Schee, P.~Romatschke, and S.~Pratt, {\it {Fully Dynamical Simulation
  of Central Nuclear Collisions}},  {\em Phys. Rev. Lett.} {\bf 111} (2013),
  no.~22 222302, [\href{http://arxiv.org/abs/1307.2539}{{\tt
  arXiv:1307.2539}}].

\bibitem{Romatschke:2015gxa}
P.~Romatschke, {\it {Light-Heavy Ion Collisions: A window into pre-equilibrium
  QCD dynamics?}},  {\em Eur. Phys. J.} {\bf C75} (2015), no.~7 305,
  [\href{http://arxiv.org/abs/1502.04745}{{\tt arXiv:1502.04745}}].

\bibitem{Weller:2017tsr}
R.~D. Weller and P.~Romatschke, {\it {One fluid to rule them all: viscous
  hydrodynamic description of event-by-event central p+p, p+Pb and Pb+Pb
  collisions at $\sqrt{s}=5.02$ TeV}},
  \href{http://arxiv.org/abs/1701.07145}{{\tt arXiv:1701.07145}}.

\bibitem{Romatschke:2009kr}
P.~Romatschke, {\it {Relativistic Viscous Fluid Dynamics and Non-Equilibrium
  Entropy}},  {\em Class. Quant. Grav.} {\bf 27} (2010) 025006,
  [\href{http://arxiv.org/abs/0906.4787}{{\tt arXiv:0906.4787}}].

\bibitem{Grozdanov:2015kqa}
S.~Grozdanov and N.~Kaplis, {\it {Constructing higher-order hydrodynamics: The
  third order}},  {\em Phys. Rev.} {\bf D93} (2016), no.~6 066012,
  [\href{http://arxiv.org/abs/1507.02461}{{\tt arXiv:1507.02461}}].

\bibitem{Heller:2013fn}
M.~P. Heller, R.~A. Janik, and P.~Witaszczyk, {\it {Hydrodynamic Gradient
  Expansion in Gauge Theory Plasmas}},  {\em Phys. Rev. Lett.} {\bf 110}
  (2013), no.~21 211602, [\href{http://arxiv.org/abs/1302.0697}{{\tt
  arXiv:1302.0697}}].

\bibitem{Florkowski:2016zsi}
W.~Florkowski, R.~Ryblewski, and M.~Spaliński, {\it {Gradient expansion for
  anisotropic hydrodynamics}},  {\em Phys. Rev.} {\bf D94} (2016), no.~11
  114025, [\href{http://arxiv.org/abs/1608.07558}{{\tt arXiv:1608.07558}}].

\bibitem{Ryu:2015vwa}
S.~Ryu, J.~F. Paquet, C.~Shen, G.~S. Denicol, B.~Schenke, S.~Jeon, and C.~Gale,
  {\it {Importance of the Bulk Viscosity of QCD in Ultrarelativistic Heavy-Ion
  Collisions}},  {\em Phys. Rev. Lett.} {\bf 115} (2015), no.~13 132301,
  [\href{http://arxiv.org/abs/1502.01675}{{\tt arXiv:1502.01675}}].

\bibitem{Habich:2015rtj}
M.~Habich, G.~A. Miller, P.~Romatschke, and W.~Xiang, {\it {Testing
  hydrodynamic descriptions of p+p collisions at $\sqrt{s}=7$ TeV}},  {\em Eur.
  Phys. J.} {\bf C76} (2016), no.~7 408,
  [\href{http://arxiv.org/abs/1512.05354}{{\tt arXiv:1512.05354}}].

\bibitem{Schukraft:2017nbn}
J.~Schukraft, {\it {QM2017: Status and Key open Questions in Ultra-Relativistic
  Heavy-Ion Physics}},  {\em Nucl. Phys.} {\bf A967} (2017) 1--10,
  [\href{http://arxiv.org/abs/1705.02646}{{\tt arXiv:1705.02646}}].

\bibitem{Nagle:2017sjv}
J.~L. Nagle, R.~Belmont, K.~Hill, J.~Orjuela~Koop, D.~V. Perepelitsa, P.~Yin,
  Z.-W. Lin, and D.~McGlinchey, {\it {Are minimal conditions for collectivity
  met in $e^+e^-$ collisions?}},  \href{http://arxiv.org/abs/1707.02307}{{\tt
  arXiv:1707.02307}}.

\bibitem{Zimdahl:1996ka}
W.~Zimdahl, {\it {Bulk viscous cosmology}},  {\em Phys. Rev.} {\bf D53} (1996)
  5483--5493, [\href{http://arxiv.org/abs/astro-ph/9601189}{{\tt
  astro-ph/9601189}}].

\bibitem{Gagnon:2011id}
J.-S. Gagnon and J.~Lesgourgues, {\it {Dark goo: Bulk viscosity as an
  alternative to dark energy}},  {\em JCAP} {\bf 1109} (2011) 026,
  [\href{http://arxiv.org/abs/1107.1503}{{\tt arXiv:1107.1503}}].

\bibitem{Blas:2015tla}
D.~Blas, S.~Floerchinger, M.~Garny, N.~Tetradis, and U.~A. Wiedemann, {\it
  {Large scale structure from viscous dark matter}},  {\em JCAP} {\bf 1511}
  (2015) 049, [\href{http://arxiv.org/abs/1507.06665}{{\tt arXiv:1507.06665}}].

\bibitem{Florkowski:2017jnz}
W.~Florkowski, E.~Maksymiuk, and R.~Ryblewski, {\it {Coupled kinetic equations
  for quarks and gluons in the relaxation time approximation}},
  \href{http://arxiv.org/abs/1710.07095}{{\tt arXiv:1710.07095}}.

\end{thebibliography}\endgroup

\end{document}